\documentclass[preprint,aps]{revtex4}

\usepackage{epsfig}

\begin{document}

\title{Limitations of the Standard Gravitational Perfect Fluid Paradigm}

\author{Philip~D.~Mannheim,~James~G.~O'Brien~and~David~Eric~Cox}

\affiliation{Department of Physics\\ University of Connecticut\\ Storrs, CT
06269, USA
\\ {\tt 
philip.mannheim@uconn.edu,obrien@phys.uconn.edu,david.cox@uconn.edu}}

\date{April 20, 2010}

\begin{abstract}
We show that the standard perfect fluid paradigm is not necessarily a valid description of a curved space steady state gravitational source. Simply by virtue of not being flat, curved space geometries have to possess intrinsic length scales, and such length scales can affect the fluid structure. For modes of wavelength of order or greater than such scales eikonalized geometrical optics cannot apply and rays are not geodesic. A set of wave mode rays that would all be geodesic in flat space (where there are no intrinsic length scales) and form a flat space perfect fluid would not all remain geodesic or of the perfect fluid form when the system is covariantized to curved space. Covariantizing thus entails not only the replacing of flat space functions by covariant ones, but also the introduction of intrinsic scales that were absent in flat space. In principle it is thus unreliable to construct  the curved space energy-momentum tensor as the covariant generalization of a geodesic-based flat spacetime energy-momentum tensor. By constructing the partition function as an incoherent average over a complete set of modes of a scalar field propagating in a curved space background, we show that for the specific case of a static, spherically symmetric geometry, the steady state energy-momentum tensor that ensues will in general be of the form $T_{\mu\nu}=(\rho+p)U_{\mu}U_{\nu}+pg_{\mu\nu}+\pi_{\mu\nu}$ where the anisotropic $\pi_{\mu\nu}$ is a symmetric, traceless rank two tensor which obeys $U^{\mu}\pi_{\mu\nu}=0$. Such a $\pi_{\mu\nu}$ type term is absent for an incoherently averaged steady state fluid in a spacetime where there are no intrinsic length scales, and in principle would thus be missed in a covariantizing of a flat spacetime $T_{\mu\nu}$. While the significance of such $\pi_{\mu\nu}$ type terms would need to be evaluated on a case by case basis, through the use of kinetic theory we reassuringly find that the effect of such $\pi_{\mu\nu}$ type terms is small for weak gravity stars where perfect fluid sources are  commonly used.

\end{abstract}

\maketitle

\section{Introduction}
\label{s1}

In gravitational theory it is standard to take the macroscopic energy-momentum tensor of a steady state gravitational  fluid source to be in the form of a perfect fluid. At the microscopic level a steady state fluid consists of a set of particles or waves whose contributions to the energy-momentum tensor are added incoherently (equal a priori probability for individual microstates). A steady state fluid thus consists of a set of particles that are only coupled to the background gravitational field but not to each other, or a set of wave modes that are only coupled to the background gravitational field but not to each other. As such, a steady state fluid is thus a fluid whose partition function is diagonal in a complete basis of particles or a complete basis of waves. In flat spacetime with metric $\eta_{\mu\nu}={\rm diag}(-1,1,1,1)$ the macroscopic energy-momentum tensor constructed as the partition function average of the microscopic energy-momentum tensor of the basis states of such a steady state fluid will have the perfect fluid form 
\begin{equation}
T_{\mu\nu}({\rm flat})=(\rho_f+p_f)U_{\mu}U_{\nu}+p_f\eta_{\mu\nu},
\label{C1}
\end{equation}
where $\rho_f$ is the fluid energy density, $p_f$ is the  fluid pressure,  and $U^{\mu}$ is a timelike velocity vector that obeys $\eta_{\mu\nu}U^{\mu}U^{\nu}=-1$. It is conventional to covariantize this expression in order to obtain the macroscopic energy-momentum tensor that is to describe a steady state fluid in a curved geometry with general metric $g_{\mu\nu}$. The covariantization prescription is to replace all flat space tensors by curved space ones and all flat space derivatives by covariant ones, so that a curved space perfect fluid is to be described by (\ref{C1}) as written in a non-flat background, viz.
\begin{equation}
T_{\mu\nu}({\rm curved})=(\rho_f+p_f)U_{\mu}U_{\nu}+p_fg_{\mu\nu},
\label{E1}
\end{equation}
where now $g_{\mu\nu}U^{\mu}U^{\nu}=-1$, with no other terms with the requisite tensor structure being deemed relevant. Remarkably, no dynamical justification for the appropriateness of such a procedure appears to have been given in the literature, and its use is simply taken as being self-evident. 

In this paper we examine this covariantization prescription and find that there are steady state curved space systems for which the covariantization prescription does not necessarily hold or for which it is only approximate, i.e. there are curved space cases in which the construction of the macroscopic energy-momentum tensor starting from steady state microphysics can lead to departures from the perfect fluid form given in (\ref{E1}). Such cases are typically associated with curved space geometries with symmetry lower than the maximally 4-symmetric  flat space Minkowski geometry. In particular, we find that the covariantization prescription does not automatically hold in a situation where it is commonly used, namely in a static, spherically symmetric system such as a star (a geometry that is only maximally 2-symmetric). Reassuringly though, we will find that for weak gravity stars the departure from a perfect fluid form will be small, to thus justify the use of perfect fluids in such cases.

To understand why there might even be a concern about the covariantization prescription, we note that there is more to covariantization than merely replacing flat space functions by their curved space generalizations. Specifically, since the act of replacing a flat space geometry by a curved space one replaces constant metric coefficients by spacetime dependent ones, the very act of covariantizing a geometry necessitates that a geometry that possesses no intrinsic lengths scales is replaced by one that does (viz. the scale on which the metric coefficients vary). Microscopic physics in curved space thus becomes sensitive to the presence of these intrinsic length scales even though there had been no analogous sensitivity in the associated flat spacetime limit.  Modes of wavelength of order or greater than such intrinsic scales cannot be described by eikonalized geometrical optics and their associated rays cannot be the geodesic ones they would have been in the absence of curvature. (In flat space where there are no intrinsic scales, modes of any wavelength are geodesic.) Thus a set of modes that were all geodesic in flat space and could all be described by geometric optics (the ingredient that actually makes their macroscopic energy-momentum tensor be of the perfect fluid form in the first place) could not all continue to obey geometric optics after covariantization. The contribution to the curved space partition function of these long wavelength modes would then necessarily lead to a macroscopic energy-momentum tensor that is not of perfect fluid form. Since the covariantizing of a wave equation would not cause modes to become coupled to each other if they had not already been coupled in flat space, following covariantization a diagonal partition function will remain diagonal, and will thus still describe a steady state system. The only difference between the covariantized and flat space cases would be that the curved space long wavelength modes would no longer be geodesic.

To be more specific, we recall that if, as for instance noted in \cite{R3a} and \cite{R4}, we conveniently set $S(x)={\rm exp}(iT(x))$ in a typical wave equation such as the curved space scalar field  wave equation $\nabla^{\mu}\nabla_{\mu}S(x)=0$ for a massless scalar field $S(x)$, we obtain
\begin{equation}
\nabla^{\mu}T\nabla_{\mu}T-i\nabla^{\mu}\nabla_{\mu}T=0.
 \label{E33}
\end{equation}
For wavelengths that are short with respect to the typical scale of the problem on which the variation of the $\nabla^{\mu}\nabla_{\mu}T$ term is important, (\ref{E33}) reduces to $\nabla^{\mu}T\nabla_{\mu}T=0$, to yield first $\nabla^{\mu}T\nabla_{\nu}\nabla_{\mu}T=0$, and then
\begin{equation}
\nabla^{\mu}T\nabla_{\mu}\nabla_{\nu}T=0
 \label{E34}
\end{equation}
since $\nabla_{\mu}\nabla_{\nu}T=\nabla_{\nu}\nabla_{\mu}T$.
Since normals to the wavefronts obey the eikonal relation
\begin{equation}
\nabla^{\mu}T=\frac{dx^{\mu}}{dq}=k^{\mu}
 \label{E35}
\end{equation}
where $q$ measures distance along the normal and $k^{\mu}$ is the wave vector of the wave, on noting that $d/dq=(dx^{\mu}/dq)(\partial/\partial x^{\mu})$, from (\ref{E34}) we thus obtain
\begin{equation}
k^{\mu}\nabla_{\mu}k^{\nu}=\frac{d^2x^{\nu}}{dq^2}+\Gamma^{\nu}_{\mu\lambda}\frac{dx^{\mu}}{dq}\frac{dx^{\lambda}}{dq}=0.
 \label{E36}
\end{equation}
Recognizing (\ref{E36}) as the massless geodesic equation, we see that in the short wavelength limit, rays move on geodesics. Moreover, exactly the same result can be obtained (see e.g. \cite{R4}) for massive scalar fields as well, with it being a general rule that short wavelength rays are geodesic. Now while the $\nabla^{\mu}\nabla_{\mu}T$ term would vanish identically for a plane wave $e^{ik\cdot x}$ with constant $k^{\mu}$ propagating in flat spacetime, in a curved space this term would not automatically vanish, and thus long wavelength curved space wave modes would not be geodesic even though these selfsame modes were geodesic in the absence of curvature. Thus in order to avoid missing any explicit curvature-dependent effects when one goes from flat space to curved space, one should covariantize not the flat space limit of (\ref{E34}), but rather one should covariantize the flat space limit of the full (\ref{E33}) instead \cite{Q1}.

As well as have an effect on wave modes, these same curved space considerations can also effect particle motions. As for instance noted in \cite{Q2,R4}, neither the equivalence principle nor general coordinate invariance prevents the presence of explicit curvature-dependent terms in particle equations of motion even though such terms would be absent in flat spacetime. Specifically, if we consider the point particle action given in \cite{R4}, viz. 
\begin{equation}
I_T=-mc\int d\tau - \kappa\int d\tau R^{\alpha}_{\phantom{\alpha}\alpha},                                                                              
\label{D9}
\end{equation}               
its variation with respect to the particle coordinate $x^{\lambda}$ leads to the equation of motion
\begin{equation}
mc \left(
\frac{d^2x^{\lambda} }{ d\tau^2} +\Gamma^{\lambda}_{\mu \nu} 
\frac{dx^{\mu}}{d\tau}\frac{dx^{\nu } }{ d\tau} \right) 
= -\kappa R^{\alpha}_{\phantom {\alpha} \alpha}\left(
\frac{d^2x^{\lambda} }{ d\tau^2} +\Gamma^{\lambda}_{\mu \nu} 
\frac{dx^{\mu}}{d\tau}\frac{dx^{\nu } }{ d\tau} \right) 
-\kappa R^{\alpha}_{\phantom {\alpha} \alpha ;\beta} \left( g^{\lambda
\beta}+
\frac{dx^{\lambda}}{d\tau}                                                      
\frac{dx^{\beta}}{d\tau}\right). 
\label{D10}
\end{equation}
Similarly for a particle with spin vector $S^{\mu}$, Weinberg \cite{Q2} considers the equation of motion 
\begin{equation}
mc\left(\frac{d^2x^{\lambda} }{ d\tau^2} +\Gamma^{\lambda}_{\mu \nu} 
\frac{dx^{\mu}}{d\tau}\frac{dx^{\nu } }{ d\tau}\right)
=-fR^{\lambda}_{\phantom{\lambda}\mu\nu\kappa}{dx^{\mu} \over d\tau}{dx^{\nu} \over d\tau}S^{\kappa},
\label{D11}
\end{equation}
where $f$ is a constant. Both (\ref{D10}) and (\ref{D11}) are general coordinate invariant equations of motion, and in the absence of curvature both reduce to the standard flat spacetime geodesic equation 
\begin{equation}
mc\left(\frac{d^2x^{\lambda} }{ d\tau^2} +\Gamma^{\lambda}_{\mu \nu} 
\frac{dx^{\mu}}{d\tau}\frac{dx^{\nu } }{ d\tau}\right)=0
\label{D12}
\end{equation}
(as written here in a curvilinear coordinate system). Moreover, both (\ref{D10}) and (\ref{D11}) obey the equivalence principle, since no matter how curved a space might be, at any given point in the spacetime one is always able to find a coordinate system in which the Christoffel symbol term can be made to vanish \cite{Q3}.  Thus while one can descend to the flat space (\ref{D12}) starting from the curved space equations such as (\ref{D10}) or (\ref{D11}), covariantizing the flat spacetime (\ref{D12}) can never uncover the presence of any curved space terms that are absent in flat spacetime. Thus for both waves and particles flat spacetime information is not a reliable guide for extracting curved space information. However, as noted by Weinberg \cite{Q2}, if the scale on which particle parameters (or analogously wave parameters) vary is much less than the scale on which the spatial curvature varies, curvature-dependent terms will be negligible, and curved space geodesic behavior for particles and waves will still result. It is the purpose of this paper to explore what happens when such curvature-dependent terms are not negligible, and to provide a general framework for determining how important they might then potentially be.

While the above analysis applies to any individual particle mode or any individual wave mode, for the fluid sources that are commonly used in astrophysics and cosmology we need to calculate partition function averages by summing over complete sets of such particles or waves. Since we can do the summation in any basis in which the partition function is diagonal, for the purposes of this paper we shall do the summation in a wave mode basis as the needed calculations then prove to be very straightforward. Moreover, wave mode bases are of central relevance in astrophysics and cosmology since white dwarf stars are stabilized by the Pauli degeneracy of the quanta of the spin one-half Dirac field, and the cosmological radiation era is described by the coupling of gravity to the cosmic  black-body radiation. 

To understand the nature of the incoherent averaging that is required, it is instructive to first consider a free Fermi gas in flat spacetime with its familiar energy density and pressure 
\begin{equation}
\rho={1\over \pi ^2\hbar^3}\int_0^{k_{\rm F}}dk k^2(k^2+m^2)^{1/2},\ \ \ \ p={1\over 3\pi ^2\hbar^3}\int_0^{k_{\rm F}}dk {k^4\over (k^2+m^2)^{1/2}},
\label{D13}
\end{equation}
as summed incoherently over all the modes up to the Fermi momentum $k_{\rm F}$. For such a Fermi gas the energy-momentum tensor has the form 
\begin{equation}
T_{\mu\nu}({\rm gas})=(\rho+p)U_{\mu}U_{\nu}+p\eta_{\mu\nu},
\label{D14}
\end{equation}
where $U_{\mu}$ is unit length timelike vector and $\eta_{\mu\nu}$ is the Minkowski metric. For a free  massive quantum-mechanical fermionic field in flat spacetime the energy-momentum tensor of the field is given by 
\begin{equation}
T_{\mu\nu}({\rm field})=i\bar{\psi}\gamma_{\mu}\partial_{\nu}\psi.
\label{D15}
\end{equation}
When this $T_{\mu\nu}({\rm field})$ is evaluated in a fermionic quantum state with momentum $k_{\mu}=(E_k,\bar{k})$ where  $E_k=(\bar{k}^2+m^2)^{1/2}$, up to irrelevant normalization factors one obtains
\begin{equation}
T_{\mu\nu}({\rm state})={k_{\mu}k_{\nu} \over E_k}.
\label{D16}
\end{equation}
Since this expression refers to one state alone, it is coherent. However, it is not of the perfect fluid form, since for $k_{\mu}=(E_k,0,0,k)$ for instance we find that  $T_{\mu\nu}({\rm state})$ evaluates to
\begin{eqnarray}
T_{\mu\nu}(E_k,0,0,k)=\pmatrix{
E_k&0&0&k\cr
0&0&0&0 \cr
0&0&0&0 \cr
k&0&0&k^2/E_k \cr}
\label{D17}
\end{eqnarray}                                 
with the off-diagonal $T_{03}(E_k,0,0,k)$ not being zero. To eliminate this off-diagonal term we now incoherently add the contribution of the mode propagating in the negative $z$ direction, to give
\begin{eqnarray}
T_{\mu\nu}(E_k,0,0,k)+T_{\mu\nu}(E_k,0,0,-k)=\pmatrix{
2E_k&0&0&0\cr
0&0&0&0 \cr
0&0&0&0 \cr
0&0&0&2k^2/E_k \cr}
\label{D18}
\end{eqnarray}
While now diagonal this is still not of the perfect fluid form as only the $(3,3)$ component of the pressure tensor is non-zero. To bring it to a perfect fluid form we must add modes propagating in the $\pm x$ and $\pm y$ directions as well, with this six state sum then yielding 
\begin{eqnarray}
T_{\mu\nu}({\rm 6~states})=
\pmatrix{
6E_k&0&0&0\cr
0&2k^2/E_k&0&0 \cr
0&0&2k^2/E_k&0 \cr
0&0&0&2k^2/E_k \cr}.
\label{D19}
\end{eqnarray}
Finally,  when we sum over all directions and magnitudes of $\bar{k}$ up to the Fermi momentum, (\ref{D13}) and (\ref{D14}) are then recovered. 

The essence of the above calculation is that we first evaluate matrix elements of $T_{\mu\nu}$ in individual states and then add the matrix elements, rather than first add states and then evaluate the matrix elements. The summation over matrix elements in individual states is incoherent, while the summation over states first is coherent and generates interference cross-terms. Steady state systems in statistical mechanical equilibrium are thus associated with the incoherent averaging procedure, and so it is this particular averaging that we shall perform in curved space in order to determine how such curved space sums might then look. To determine how significant such curvature-dependent effects might be in any given gravitational situation one would need to construct the macroscopic energy-momentum tensor starting from microphysics in each individual case and see what ensues. In general then, the required procedure is to first find a basis in which the partition function is diagonal and from it then determine the macroscopic energy-momentum tensor as the partition function average of the microscopic energy-momentum tensor of the basis states, and whatever form the ensuing macroscopic energy-momentum tensor then takes, that is the macroscopic form one has to use. To actually show that  we do not always recover (\ref{E1}) in general, we only need to find one explicit counterexample. We thus only need to  find one appropriate choice of metric coefficients for which one can do the incoherent partition function summation over an infinite set of wave modes analytically and then fail to recover (\ref{E1}). It is precisely such a calculation that we provide in this paper. 

In the literature the notions of steady state fluid and perfect fluid are ordinarily taken to be equivalent and are used interchangeably. However, since the results of our work hinge on an in principle difference between the two concepts, it is instructive to clarify what that difference is. For a fluid to be in a steady state we understand only that we are able to find a basis in which the partition function of the system is diagonal, with the basis states being decoupled from each other. As such, this requirement is only a requirement on the structure of the basis states, and in and of itself, it is not a requirement on the form of the energy-momentum tensor, and it is not obliged to lead to the form for the energy-momentum given in (\ref{E1}). However, in the literature it is the form given in (\ref{E1}) which is referred to as the perfect fluid form, with any departures from this form (so-called imperfect fluid terms) being thought to be caused by interactions between the basis states, interactions that are associated with transport phenomena such as viscosity and heat conduction. The point of our work here is that in curved spacetime, one can obtain departures from the form given in (\ref{E1}) even when the basis modes are not in interaction with each other at all, i.e. that after relaxation to steady state in curved space one can obtain a form for the energy-momentum tensor different from that given in (\ref{E1}). This situation is to be contrasted with the situation that obtains in flat spacetime, since there it is the interactions associated with the non-vanishing of the collision integral term in the kinetic theory Boltzmann equation that lead to the viscosity and heat conduction dependent terms associated with fluids  that are not perfect. However, that wisdom does not carry over to curved space since gravitational interactions are described not by a adding a gravitational scattering contribution to the Boltzmann equation collision integral, but by treating gravity as being due to curvature instead. As we explicitly show in this paper, after relaxing to steady state in the presence of gravity, the energy-momentum tensor of a fluid need not have the form given in (\ref{E1}).

In order to ascertain for which steady state systems a departure from the perfect fluid form might be the most pronounced, it is instructive to analyze the geometric structure of two of the most commonly encountered geometries in astrophysics, the maximally 2-symmetric geometry associated with a static, spherically symmetric star, and the maximally 3-symmetric Robertson-Walker geometry associated with  a homogeneous and isotropic cosmology. We do not consider systems such as  an accretion disk but the conclusions would be analogous. 

For the case first of a star, we note that unlike the maximally 4-symmetric flat spacetime geometry in which steady state fluids do have the form given in (\ref{C1}), spherical systems have much lower (only maximally 2-symmetric) symmetry and are only isotropic about a single point (the center of the system). Such a symmetry only requires for any given tensor $A_{\mu\nu}$ that its  $A_{\theta}^{\phantom{\theta}\theta}$ and $A_{\phi}^{\phantom{\phi}\phi}$ components be equal, and imposes no restriction on $A_{r}^{\phantom{r}r}$. A familiar example of this is the form of the Einstein tensor $G_{\mu\nu}$ in the static, spherical geometry 
\begin{equation}
ds^2=-B(r)dt^2+A(r)dr^2+r^2d\theta^2+r^2\sin^2\theta d\phi^2,
\label{E2}
\end{equation}
where the non-zero components of $G_{\mu\nu}$ are given by
\begin{eqnarray}
G_{00}&=&-\frac{B}{r^2}+\frac{B}{r^2A}-\frac{BA^{\prime}}{rA^2},
\nonumber\\
G_{rr}&=&\frac{A}{r^2}-\frac{1}{r^2}-\frac{B^{\prime}}{rB},
\nonumber\\
G_{\theta\theta}&=&\frac{G_{\phi\phi}}{\sin^2\theta} =-\frac{r^2B^{\prime\prime}}{2AB}+\frac{r^2A^{\prime}B^{\prime}}{4A^2B}+\frac{r^2B^{\prime 2}}{4AB^2}-\frac{rB^{\prime}}{2AB}+\frac{rA^{\prime}}{2A^2}.
\label{E3}
\end{eqnarray}
As we see, there no relation of the form $G_{r}^{\phantom{r}r}=G_{\theta}^{\phantom{\theta}\theta}$ in (\ref{E3}), and not only that, there could not be since the radial component of the Bianchi identity, viz. $G^{r\nu}_{\phantom{\mu\nu};\nu}=\partial_rG^{rr}+(A^{\prime}/A+B^{\prime}/2B+2/r)G^{rr}-(r/A)G^{\theta\theta}-(r{\rm sin}^2\theta /A)G^{\phi\phi}+(B^{\prime}/2A)G^{00}=0$, relates the radial derivative of $G^{rr}$ to $G^{\theta\theta}$, to thereby force $G^{rr}$ to be a first-order derivative function of  $r$ and $G^{\theta\theta}$ to be a second-order one. 

This same problem is not evaded if one works in isotropic coordinates, where, because of coordinate invariance, the metric can actually  be brought to  a form
\begin{equation}
ds^2=-H(\rho)dt^2+J(\rho)(d\rho^2+\rho^2d\theta^2+\rho^2\sin^2\theta d\phi^2)
\label{E4}
\end{equation}
that does have the generic perfect fluid form $g_{\rho}^{\phantom{\rho}\rho}=g_{\theta}^{\phantom{\theta}\theta}=g_{\phi}^{\phantom{\phi}\phi}$. Specifically, even in this coordinate system the Einstein tensor 
\begin{eqnarray}
G_{00}&=&\frac{2HJ^{\prime}}{\rho J^2}-\frac{3HJ^{\prime 2}}{4 J^3}+\frac{HJ^{\prime \prime}}{J^2},\nonumber\\
G_{\rho\rho}&=&-\frac{J^{\prime}}{\rho J}-\frac{H^{\prime}J^{\prime}}{2H J}-\frac{J^{\prime 2}}{4J^2}-\frac{H^{\prime}}{\rho H}
\nonumber\\
G_{\theta\theta}&=&\frac{G_{\phi\phi}}{\sin^2\theta} =-\frac{\rho^2J^{\prime \prime}}{ 2J}-\frac{\rho J^{\prime}}{ 2J}+\frac{\rho^2J^{\prime 2}}{ 2J^2}-\frac{\rho H^{\prime}}{ 2H}+\frac{\rho^2H^{\prime 2}}{ 4H^2}-\frac{\rho^2 H^{\prime \prime}}{ 2H}
\label{E5}
\end{eqnarray}
is still not of a perfect fluid form, and indeed must still not be since now the Bianchi identity relates the radial derivative of $G^{\rho\rho}$ to $G^{\theta\theta}$. Since the gravitational equations of motion would relate gravitational tensors such as the Einstein tensor  to the energy-momentum tensor, even in isotropic coordinates we should not expect the components of $T_{\mu\nu}$ to necessarily obey $T_{\rho}^{\phantom{\rho}\rho}=T_{\theta}^{\phantom{\theta}\theta}=T_{\phi}^{\phantom{\phi}\phi}$. We thus anticipate, and shall indeed find, that for a steady state star there will be in principle departures from the perfect fluid form.

While the spatial symmetry of a static, spherically symmetric star is quite low (the geometry is only isotropic about one single point), in the Robertson-Walker case the symmetry of the spatial geometry is very high (the geometry is isotropic about every spatial point), being in fact as high as the spatial symmetry of flat spacetime.  Thus just as the incoherent averaging over all of the modes of classical massless scalar or classical Maxwell fields propagating in a Minkowski background geometry yields a perfect fluid energy-momentum tensor \cite{R1,R2,R3}, incoherent averaging of the modes of the same fields in a background Robertson-Walker geometry does so as well \cite{R1,R2,R3}. While it is the case that a Robertson-Walker geometry does possess an intrinsic length scale (viz. that associated with the spatial 3-curvature $k$), because of the high spatial symmetry, the only role of this intrinsic scale is to affect how the energy density and pressure depend on temperature \cite{R3} but not to generate any imperfect fluid terms in the energy-momentum tensor. The effect which we identify in this paper could thus only be manifest in geometries with a spatial symmetry lower than that which obtains for a Robertson-Walker geometry.

In Sec.~\ref{s2} of this paper we describe the incoherent averaging procedure in detail. In Sec.~\ref{s3} we apply it to a fluid in a specific, exactly solvable, static, spherically symmetric geometry, to find that we do not obtain a perfect fluid form. In Sec.~\ref{s4} we examine the tensor structure of the energy-momentum tensor that we obtain by incoherent averaging, and identify the presence of a non-perfect fluid $\pi_{\mu\nu}$ term in it. In Sec.~\ref{s5} we show that through the imposition of boundary conditions  there can be departures from a perfect fluid form for finite-sized systems even in flat spacetime. Finally, in Sec.~\ref{s6} we discuss the implications of kinetic theory for the structure of fluids, and show that even for Newtonian gravity there are in principle differences between stars and clusters of galaxies, even though both systems are static and spherically symmetric when in steady state. Such differences would need to be taken into consideration when these particular systems become relativistic. Reassuringly though, through the use of this same kinetic theory analysis we are able to provide a justification for the use of perfect fluids sources in weak gravity stars, though any other gravitational situations  would need to be analyzed on a case by case basis.

\section{The Curved Space Energy-Momentum Tensor}
\label{s2}

For our purposes here an appropriate model to study is a minimally coupled massless scalar field $S(x)$ with action 
\begin{equation}
I=-\int d^4 x (-g)^{1/2} \frac{1}{2}\nabla_{\mu}S\nabla^{\mu}S
\label{E6}
\end{equation}
propagating in the background associated with the metric of (\ref{E4}). For the scalar field the equation of motion is given by $\nabla_{\mu}\nabla^{\mu}S=0$, i.e. by
\begin{equation}
-\frac{1}{H}\frac{\partial^2 S}{\partial t^2}
+\frac{1}{H^{1/2}J^{3/2}\rho^2}\frac{\partial}{\partial \rho}\left[H^{1/2}J^{1/2}\rho^2\frac{\partial S}{\partial \rho}\right]
+\frac{1}{J\rho^2}\left[\frac{1}{\sin \theta}\frac{\partial}{\partial \theta}\left(\sin \theta\frac{\partial S}{\partial \theta}\right)+
\frac{1}{\sin^2\theta}\frac{\partial^2 S}{\partial \phi^2}\right]
=0,
\label{E7}
\end{equation}
and the energy-momentum tensor is of the form
\begin{equation}
T_{\mu\nu}=\nabla_{\mu}S\nabla_{\nu}S-\frac{1}{2}g_{\mu\nu}\nabla_{\alpha}S\nabla^{\alpha}S.
\label{E8}
\end{equation}
The equation of motion can be separated, and for a real scalar field the general solution can be expressed in terms of four characteristic solutions 
\begin{eqnarray}
&&S_1(x)=N_{\ell}^m\sin(\omega t)S_{\omega,\ell}(\rho)P_{\ell}^m(\theta)\sin(m\phi),
\qquad S_2(x)=N_{\ell}^m\sin(\omega t)S_{\omega,\ell}(\rho)P_{\ell}^m(\theta)\cos(m\phi),
\nonumber\\
&&S_3(x)=N_{\ell}^m\cos(\omega t)S_{\omega,\ell}(\rho)P_{\ell}^m(\theta)\sin(m\phi),
\qquad S_4(x)=N_{\ell}^m\cos(\omega t)S_{\omega,\ell}(\rho)P_{\ell}^m(\theta)\cos(m\phi),
\nonumber\\
\label{E9}
\end{eqnarray}
where $N_{\ell}^m$ is given by 
\begin{equation}
N_{\ell}^m=(-1)^m\left[\frac{(2\ell+1)(\ell-m)!}{4\pi (\ell+m)!}\right]^{1/2},
\label{E10}
\end{equation}
and where the radial term obeys 
\begin{equation}
\left[\frac{\omega^2}{H}
+\frac{1}{H^{1/2}J^{3/2}\rho^2}\frac{d}{d \rho}\left(H^{1/2}J^{1/2}\rho^2\frac{d }{d\rho}\right)-\frac{\ell(\ell+1)}{J\rho^2}\right]S_{\omega,\ell}(\rho)=0.
\label{E11}
\end{equation}

For the incoherent averaging procedure in the static, spherically symmetric case, we note that in a given mode such as $S_1(x)$ a quantity such as $\nabla_{\alpha}S\nabla^{\alpha}S$ evaluates to
\begin{eqnarray}
\nabla_{\alpha}S_1\nabla^{\alpha}S_1&=&-\frac{\omega^2}{H}\cos^2(\omega t)[S_{\omega,\ell}N_{\ell}^mP_{\ell}^m]^2\sin^2(m\phi)
+\frac{1}{J}\sin^2(\omega t)\left[\frac{d S_{\omega,\ell}}{d \rho}\right]^2[N_{\ell}^mP_{\ell}^m]^2\sin^2(m\phi)
\nonumber\\
&+&\frac{1}{J\rho^2}\sin^2(\omega t)[S_{\omega,\ell}]^2\left[[N_{\ell}^m]^2\bigg{[}\frac{d P_{\ell}^{m}}{d \theta}\right]^2\sin^2(m\phi)
+\frac{m^2}{\sin^2\theta}[N_{\ell}^mP_{\ell}^{m}]^2\cos^2(m\phi)\bigg{].}
\label{E12}
\end{eqnarray}
On now adding to this expression those obtained when one uses the  other three above solutions, one obtains
\begin{eqnarray}
\sum_{i=1}^{i=4}\nabla_{\alpha}S_i\nabla^{\alpha}S_i &=&-\frac{\omega^2}{H}[S_{\omega,\ell}N_{\ell}^mP_{\ell}^m]^2
+\frac{1}{J}\left[\frac{d S_{\omega,\ell}}{d \rho}\right]^2[N_{\ell}^mP_{\ell}^m]^2
\nonumber\\
&&+\frac{1}{J\rho^2}[S_{\omega,\ell}]^2\left[[N_{\ell}^m]^2\left[\frac{d P_{\ell}^{m}}{d \theta}\right]^2
+\frac{m^2}{\sin^2\theta}[N_{\ell}^mP_{\ell}^{m}]^2\right].
\label{E13}
\end{eqnarray}
Using standard properties of the spherical harmonics, 
\begin{eqnarray}
\sum_m [N_{\ell}^m]^2[P_{\ell}^m]^2&=&\frac{(2\ell+1)}{4\pi},
\nonumber\\
\sum_m [N_{\ell}^m]^2m^2[P_{\ell}^m]^2&=&\frac{(2\ell+1)\ell(\ell+1)\sin^2\theta}{8\pi},
\nonumber\\
\sum_m [N_{\ell}^m]^2\left[\frac{dP_{\ell}^m}{d\theta}\right]^2&=&\frac{(2\ell+1)\ell(\ell+1)}{8\pi},
\label{E14}
\end{eqnarray}
one sums over the azimuthal quantum number $m$ just as in \cite{R1},  to obtain
\begin{equation}
\sum_m\sum_{i=1}^{i=4} \nabla_{\alpha}S_i\nabla^{\alpha}S_i=\frac{(2\ell+1)}{4\pi}\left[\left(-\frac{\omega^2}{H}
+\frac{\ell(\ell+1)}{J\rho^2}\right)[S_{\omega,\ell}]^2+\frac{1}{J} \left[\frac{dS_{\omega,\ell}}{d\rho}\right]^2\right]=K(\omega,\rho,\ell),
\label{E15}
\end{equation}
with (\ref{E15}) serving to define $K(\omega,\rho,\ell)$. With regard to (\ref{E15}), we note that because the angular part of the metric is maximally 2-symmetric, the sum on the azimuthal $m$ removes any dependence on the angle $\theta$. Repeating this same procedure for the rest of $T_{\mu\nu}$ of (\ref{E8}) then yields (again following \cite{R1})
\begin{eqnarray}
 T_{00}(\omega,\rho,\ell)&=&\frac{(2\ell+1)}{4\pi}\omega^2[S_{\omega,\ell}]^2+\frac{HK}{2},
 \nonumber \\
 T_{\rho\rho}(\omega,\rho,\ell)&=&\frac{(2\ell+1)}{4\pi}\left[\frac{d S_{\omega,\ell}}{d \rho}\right]^2-\frac{JK}{2},
 \nonumber\\
 T_{\theta\theta}(\omega,\rho,\ell)&=&\frac{1}{\sin^2\theta}T_{\phi\phi}(\omega,\rho,\ell)=
 \frac{(2\ell+1)\ell(\ell+1)}{8\pi}[S_{\omega,\ell}]^2-\frac{\rho^2JK}{2},
 \label{E16}
\end{eqnarray}
with all other components of $T_{\mu\nu}$ being zero.

To complete the incoherent averaging we need to sum over all $\ell$ values as well, an infinite summation.  However, if we revert back to flat space where $H(\rho)=J(\rho)=1$, the radial equation is then solved by the spherical Bessel functions $j_{\ell}(\omega\rho)$, and the sum on $\ell$ can then be performed analytically using the completeness relations for Bessel functions:
\begin{eqnarray}
&&\sum_{\ell} (2\ell+1)j_\ell^2=1,\qquad \sum_\ell (2\ell+1)j_{\ell}\frac{dj_{\ell}}{d\rho}=0,\qquad \sum_{\ell} (2\ell+1)\left[\frac{dj_{\ell}}{d\rho}\right]^2=\frac{\omega^2}{3},
\nonumber\\
&&\sum_{\ell} (2\ell+1)j_{\ell}\frac{d^2j_{\ell}}{d\rho^2}=-\frac{\omega^2}{3},\qquad \sum_{\ell} (2\ell+1)\ell(\ell+1)j_{\ell}^2=\frac{2\omega^2\rho^2}{3}.
\label{E17}
\end{eqnarray}
On defining $K(\omega,\rho)=\sum_{\ell}K(\omega,\rho,\ell)$ and $T_{\mu\nu}(\omega,\rho)=\sum_{\ell}T_{\mu\nu}(\omega,\rho,\ell)$, one obtains $K(\omega,\rho)=0$ and 
\begin{equation}
 T_{00}(\omega,\rho)=\frac{\omega^2}{4\pi},\qquad 
 T_{\rho\rho}(\omega,\rho)=\frac{\omega^2}{12\pi},\qquad T_{\theta\theta}(\omega,\rho)=\frac{1}{\sin^2\theta}T_{\phi\phi}(\omega,\rho)=\frac{\omega^2\rho^2}{12\pi},
 \label{E18}
\end{equation}
One thus obtains none other than the flat space perfect fluid form given in (\ref{C1}), with the averaging on $\ell$ removing any dependence on the coordinate $\rho$ from $T_{\mu}^{\phantom{\mu}\nu}$ because of the maximal 3-symmetry of the spatial part of a Minkowski metric. The incoherent averaging prescription thus gives a perfect fluid in flat spacetime, just as one would want.

\section{Incoherent Averaging in Curved Spacetime}
\label{s3}

To follow this same procedure in curved space is not at all as straightforward, since for an arbitrary choice of $H(\rho)$ and $J(\rho)$ the radial equation will not necessarily be solvable in terms of named functions, and the needed completeness relation for the ensuing modes may not even be known at all. Moreover, in the dynamical case where the energy-momentum tensor is used as the source of tensors such as the Einstein tensor given in (\ref{E5}), one has to solve for $H(\rho)$ and $J(\rho)$ self-consistently, a procedure that would not only not appear at all likely to yield an analytic result, since it involves an infinite summation, it does not immediately lend itself to numerical approximation either. However, to test the viability of the perfect fluid assumption itself, one only has to seek an appropriate choice of $H(\rho)$ and $J(\rho)$ for which one can do the $\ell$ summation analytically, to see whether or not the relation $T_{\rho\rho}=T_{\theta\theta}/\rho^2$ then does in fact ensue. For our purposes here we do not need to impose the Einstein equations since it is not the gravitational equations of motion that cause the fluid to be perfect. Rather, one already takes a steady state  energy-momentum tensor to be of perfect fluid form in flat spacetime before gravity is even introduced. Hence we only need to ask whether or not one obtains a perfect fluid when one repeats the above flat spacetime averaging calculation in a given external gravitational field. 

Since we are able to do the $\ell$ summation analytically for Bessel functions, in the curved space case we shall seek a form for $H(\rho)$ and $J(\rho)$ for which the radial equation can be reduced to the Bessel equation. To this end we set $J(\rho)=H(\rho)=f^2(\rho)$. For this choice, the metric in (\ref{E4}) reduces to
\begin{equation}
ds^2=f^2(\rho)[-dt^2+d\rho^2+\rho^2d\theta^2+\rho^2\sin^2\theta d\phi^2],
\label{E19}
\end{equation}
to thus be conformal to flat. However, it is not flat since in it the 
Einstein tensor in (\ref{E5}) takes the non-vanishing form 
\begin{eqnarray}
G_{00}&=&\frac{4f^{\prime}}{\rho f}-\frac{f^{\prime 2}}{f^2}+\frac{2f^{\prime \prime}}{f},~~~
G_{\rho\rho}=-\frac{4f^{\prime}}{\rho f}-\frac{3f^{\prime 2}}{f^2},
\nonumber \\
G_{\theta\theta}&=&\frac{G_{\phi\phi}}{\sin^2\theta} =-\frac{2\rho f^{\prime}}{f}+\frac{\rho^2f^{\prime 2}}{f^2}-\frac{2\rho^2f^{\prime \prime}}{f},
\label{E20}
\end{eqnarray}
a form which is still not a perfect fluid unless $f(\rho)$  just happens to obey
\begin{equation}
\frac{f^{\prime\prime}}{f}-\frac{f^{\prime}}{\rho f}-\frac{2f^{\prime 2}}{f^2}=0.
\label{E21}
\end{equation}

For the metric of (\ref{E19}), on substituting  $S_{\omega,\ell}(\rho)=Q_{\omega,\ell}(\rho)/f(\rho)$ the radial equation (\ref{E11}) reduces to
\begin{equation}
\left[\omega^2
+\frac{d^2}{d \rho^2}+2\frac{d}{d\rho}-\frac{\ell(\ell+1)}{\rho^2}-\frac{1}{f}\left(f^{\prime\prime}+\frac{2f^{\prime}}{\rho}\right)\right]Q_{\omega,\ell}(\rho)=0.
\label{E22}
\end{equation}
Thus, for any choice of $f(\rho)$ for which
\begin{equation}
-\frac{1}{f}\left(f^{\prime\prime}+\frac{2f^{\prime}}{\rho}\right)=\kappa^2
\label{E23}
\end{equation}
where $\kappa^2$ is positive, one finds that (\ref{E22}) then reduces to none other than the Bessel function equation
\begin{equation}
\left[\omega^2+\kappa^2
+\frac{d^2}{d \rho^2}+\frac{2}{\rho}\frac{d}{d \rho}-\frac{\ell(\ell+1)}{\rho^2}\right]Q_{\omega,\ell}(\rho)=0,
\label{E24}
\end{equation}
with immediate solution $Q_{\omega,\ell}(\rho)=j_{\ell}(\lambda \rho)$ where $\lambda=(\omega^2+\kappa^2)^{1/2}$. For (\ref{E23}) solutions are readily given as 
\begin{equation}
f(\rho)=\frac{[\alpha\sin(\kappa\rho)+\beta\cos(\kappa\rho)]}{\rho},
\label{E25}
\end{equation}
and thus form a whole family of solutions labeled by all real values of the parameters $\kappa$, $\alpha$ and $\beta$. For any $f(\rho)$ which obeys (\ref{E23}), the Einstein tensor reduces to 
\begin{eqnarray}
G_{00}&=&-\frac{f^{\prime 2}}{f^2}-2\kappa^2,~~~
G_{\rho\rho}=-\frac{4f^{\prime}}{\rho f}-\frac{3f^{\prime 2}}{f^2},
\nonumber \\
G_{\theta\theta}&=&\frac{G_{\phi\phi}}{\sin^2\theta} =\frac{2\rho f^{\prime}}{f}+\frac{\rho^2f^{\prime }}{f}+2\rho^2\kappa^2,
\label{E26}
\end{eqnarray}
and in solutions of the form given in (\ref{E25}) is still not in the form of a perfect fluid, with the solutions in (\ref{E25}) not obeying (\ref{E21}).

To now evaluate the energy-momentum tensor when (\ref{E23}) is imposed, on summing over $\ell$ as before, one obtains
\begin{equation}
K(\omega,\rho)=\frac{(f^{\prime 2}+\kappa^2)}{4\pi f^4},
\label{E27}
\end{equation}
with the incoherently-averaged energy-momentum tensor itself being given by
\begin{eqnarray}
 T_{00}(\omega,\rho)&=&\frac{1}{4\pi}\left[\frac{\omega^2}{f^2} +\frac{f^{\prime 2}}{2f^4}+\frac{\kappa^{2}}{2f^2}\right],
 \nonumber \\
 T_{\rho\rho}(\omega,\rho)&=&\frac{1}{4\pi}\left[\frac{\omega^2}{3f^2} +\frac{f^{\prime 2}}{2f^4}-\frac{\kappa^{2}}{6f^2}\right],
 \nonumber\\
 T_{\theta\theta}(\omega,\rho)&=&\frac{T_{\phi\phi}}{\sin^2\theta} =\frac{1}{4\pi}\left[\frac{\rho^2\omega^2}{3f^2} -\frac{\rho^2f^{\prime 2}}{2f^4}-\frac{\rho^2\kappa^{2}}{6f^2}\right].
 \label{E28}
\end{eqnarray}
As we see, $T_{\mu\nu}(\omega,\rho)$ is not in the form of a perfect fluid since $T_{\rho}^{\phantom{\rho}\rho}- T_{\theta}^{\phantom{\theta}\theta}=f^{\prime 2}/4\pi f^6$ is not zero. Since our analysis is valid  for any $f(\rho)$ which obeys (\ref{E25}), we recognize a whole family of metrics for which a perfect fluid is not obtained. In general then, the use of perfect fluid sources for spherically symmetric gravitational systems must be regarded as open to question. 

The essence of the calculation that we have presented here is that we start with a free flat spacetime scalar field that possess kinetic energy but no potential energy. In the absence of any potential energy there are no scalar field interaction terms (such as $S^3$ or $S^4$ etc.), and both the wave equation for the scalar field and the associated energy-momentum tensor can be diagonalized in a complete set of flat spacetime plane wave basis modes, with the system thus being in a steady state. An incoherent averaging over these basis modes leads to a flat spacetime energy-momentum tensor that has the form of a perfect fluid (c.f. (\ref{E18})). We then extend this same scalar field theory to curved space. The system continues to be in steady state since both the curved space wave equation and energy-momentum tensor are diagonal in the curved space scalar field wave mode basis. On incoherently summing over the curved space mode basis we obtain an energy-momentum tensor (c.f. (\ref{E28})) that is not the covariant generalization of the flat spacetime energy-momentum tensor which had been constructed by the same procedure. Covariantizing the wave equation and then summing over modes thus gives a different energy-momentum tensor than first summing over the modes in flat spacetime and then covariantizing the energy-momentum tensor that ensues. It is because of this mismatch (due to the drop in symmetry from maximally 4-symmetric to only maximally 2-symmetric and the concomitant appearance of intrinsic length scales), that covariantizing the form of a flat space energy-momentum tensor can be misleading. We thus recognize that in covariantizing a system one should covariantize not the incoherently averaged flat spacetime energy-momentum tensor, but rather the flat spacetime steady state basis modes themselves, and whatever incoherently averaged curved space energy-momentum tensor then ensues, that is the correct one for the problem. Thus  it does not follow  that if a steady state flat spacetime fluid is a perfect  fluid then it will remain so in the presence of curvature. However, if matter field modes are in steady state in flat spacetime, the matter field modes themselves will still remain in steady state when curvature is introduced.

\section{The Tensor Structure of the Curved Space Fluid}
\label{s4}

To characterize the tensor structure of the energy-momentum tensor that the curved space incoherent averaging procedure has led us to, we recall \cite{R3aa,R3a} that in terms of a timelike vector $U_{\mu}$, the ten independent components of a general curved space fluid energy-momentum tensor can be written in the form
\begin{equation}
T_{\mu\nu}=\rho_fU_{\mu}U_{\nu}+p_f(U_{\mu}U_{\nu}+g_{\mu\nu})+\pi_{\mu\nu}+q_{\mu}U_{\nu}+q_{\nu}U_{\mu},
\label{E29}
\end{equation}
where $U_{\mu}U_{\nu}+g_{\mu\nu}$, $q_{\mu}$, and $\pi_{\mu\nu}$ obey $U^{\mu}(U_{\mu}U_{\nu}+g_{\mu\nu})=0$, $U^{\mu}q_{\mu}=0$, $\pi_{\mu\nu}=\pi_{\nu\mu}$, $g^{\mu\nu}\pi_{\mu\nu}=0$, $U^{\mu}\pi_{\mu\nu}=0$. With there being only three independent components in (\ref{E28}), for our purposes here we have no need for the $q_{\mu}U_{\nu}+q_{\nu}U_{\mu}$ term (a term that in kinetic theory is usually identified with heat transfer), but we do need the anisotropic $\pi_{\mu\nu}$. In terms of the general metric of (\ref{E4}), and with $U_{\mu}=(H^{1/2},0,0,0)$ as usual, the non-zero elements of $\pi_{\mu\nu}$ and $T_{\mu\nu}$ of (\ref{E29}) can be written as 
\begin{eqnarray}
\pi_{\rho\rho}&=&2q_fJ,\qquad \pi_{\theta\theta}=-\rho^2Jq_f,\qquad \pi_{\phi\phi}=-\rho^2Jq_f\sin^2\theta,
\nonumber\\
T_{00}&=&H\rho_f,\qquad T_{\rho\rho}=J(p_f+2q_f),\qquad T_{\theta\theta}=T_{\phi\phi}/\sin^2\theta=\rho^2J(p_f-q_f),
\label{E29a}
\end{eqnarray}
where just like $\rho_f$ and $p_f$, the conveniently introduced quantity $q_f$ is equally a general coordinate scalar.  For the energy-momentum tensor of (\ref{E28}) and the metric of (\ref{E19}) we thus identify
\begin{equation}
 \rho_f=\frac{1}{4\pi}\left[\frac{\omega^2}{f^4} +\frac{f^{\prime 2}}{2f^6}+\frac{\kappa^{2}}{2f^4}\right],\qquad
 p_f=\frac{1}{4\pi}\left[\frac{\omega^2}{3f^4} -\frac{f^{\prime 2}}{6f^6}-\frac{\kappa^{2}}{6f^4}\right],\qquad
q_f=
\frac{f^{\prime 2}}{12\pi f^6},
 \label{E30}
\end{equation}
with a summation over $\omega$ being implicit. The energy-momentum tensor that we have constructed here by incoherent averaging thus nicely falls into the general class of fluid energy-momentum tensors given in \cite{R3aa} and \cite{R3a}.

We would like to stress here that while the anisotropic $\pi_{\mu\nu}$ type term would be associated with viscosity effects in flat spacetime physics, the generic decomposition of the energy-momentum tensor as given in (\ref{E29}) requires no such identification. Specifically, the decomposition of (\ref{E29}) is a purely kinematic one involving tensors and vectors which are constructed solely so as to be transverse to the fluid velocity $U_{\mu}$, with (\ref{E29}) giving the most general allowed form containing such quantities. Independent of any dynamical considerations, the energy-momentum tensor must always have the form given in (\ref{E29}) (the form of (\ref{E29}) contains precisely ten independent components), and there is nothing in the structure of (\ref{E29}) which obliges a curved space $\pi_{\mu\nu}$ type term to be identified solely with non-steady state effects. While any curved space viscosity effect would of course be described by a $\pi_{\mu\nu}$ type term, nothing precludes such terms from describing steady state curved space fluids as well. The characterization of the $\pi_{\mu\nu}$ term as a viscosity term comes from experience with fluids in flat spacetime, and flat spacetime experience is not an adequate enough guide for the description of fluids in curved space. Nothing in the structure of (\ref{E29}) forbids a fully relaxed steady state fluid in curved space from possessing such an anisotropic $\pi_{\mu\nu}$, and not only that, the analysis of this paper shows that in general one should anticipate that for steady state fluids in curved space such terms can actually occur \cite{R3b}.

For the form of $T_{\mu\nu}$ given in (\ref{E29}), in the geometry associated with the metric of (\ref{E4}) the covariant conservation condition $T^{\mu\nu}_{\phantom{\mu\nu};\nu}=0$ yields
\begin{equation}
p_f^{\prime}+2q_f^{\prime}+3\left(\frac{J^{\prime}}{J}+\frac{2}{\rho}\right)q_f+\frac{H^{\prime}}{2H}(\rho_f+p_f+2q_f)=0,
 \label{E31}
\end{equation}
to thus show the explicit role played by the $q_f$ term in the energy and momentum balance.
Moreover, not only does $q_f$ act analogously to $p_f$ in the covariant conservation condition, it even does so in dynamical equations such as the Einstein equations $G_{\mu\nu}=-8\pi G T_{\mu\nu}$. Specifically, if we work in the weak gravity limit where we can set $H(\rho)=1+h(\rho)$, $J(\rho)=1+j(\rho)$ with $h(\rho)$ and $j(\rho)$ both being of order $G$, we can consistently find solutions to the Einstein equations in which $\rho_f$ is of order one and $p_f$ and $q_f$ are both of order $G$. In the weak gravity limit the Einstein equations and the conservation condition associated with the metric of (\ref{E4}) reduce to 
\begin{eqnarray}
 \frac{2j^{\prime}}{\rho}+j^{\prime\prime}&=&-8\pi G \rho_f,\qquad
-\frac{j^{\prime}}{\rho}-\frac{h^{\prime}}{\rho}=-8 \pi G (p_f+2q_f),
 \nonumber\\
 -\frac{j^{\prime \prime}}{2}-\frac{j^{\prime}}{2\rho}-\frac{h^{\prime}}{2\rho}-\frac{ h^{\prime \prime}}{2}&=&-8 \pi G (p_f-q_f),\qquad 
 p_f^{\prime}+2q_f^{\prime}+\frac{6}{\rho}q_f+\frac{h^{\prime}}{2}\rho_f=0,
 \label{E32}
\end{eqnarray}
and to order $G$ thus have a consistent weak gravity limit in which $j+h=0$. In and of themselves then, the dynamical equations do not require $q_f$ to be at least one order in $G$ smaller than $p_f$.

Now we had noted above in Sec. \ref{s1} that for wave modes it is only the short wavelength modes that obey geometrical optics, with the long wavelength modes not being geodesic. Referring now to the incoherently averaged $T_{\mu\nu}$ given in (\ref{E28}) and (\ref{E30}), we see that in the large $\omega$ limit $T_{\mu\nu}$ does indeed reduce to a perfect fluid with the $q_f$ term becoming negligible. Perfect fluids are thus to be associated with short wavelength modes alone, with departures from a perfect fluid form being expected to occur at longer wavelengths where the geometric optics approximation no longer holds and one can no longer ignore the $\nabla^{\mu}\nabla_{\mu}T$ term in (\ref{E33}). The potential importance of such long wavelength modes thus depends on the scale of spatial variation of the system of interest, a dynamical rather than a kinematical issue, and will be analyzed further in Sec.~\ref{s6} where we will reassuringly show that for normal-sized (i.e. non-compact) weak gravity stars the effect of the $q_f$ type terms is negligible. However, the relative importance of the longer wavelength modes will increase as the size of a system is decreased, and it could thus be of interest to explore what happens when a system such as star collapses to a size of order its Schwarzschild radius \cite{R4a}.

\section{Implications of Boundary Conditions}
\label{s5}

While we have discussed possible departures from a perfect fluid form in curved space, such departures can even occur in flat space due to boundary effects of finite sized systems. Such concerns do not arise in flat space when we consider plane waves, since they fill all space. However, suppose we consider a finite-sized cavity which is filled with massless scalar modes in thermal equilibrium (a spinless black body). Now it is long known that if we take the cavity to be a cubical cavity of side of length $a$ and take the modes to obey periodic boundary conditions, while we would find corrections of order $hc/kTa$ to the familiar $T_{00} \sim T^4$ black-body formula, because of the cubic symmetry these corrections would treat all components of the pressure tensor equivalently and leave the perfect fluid form $T_{xx}=T_{yy}=T_{zz}=T_{00}/3$ (i.e. $T_{rr}=T_{\theta\theta}/r^2=T_{\phi\phi}/r^2\sin^2\theta=T_{00}/3$) intact. These finite size effects would only be significant for wavelengths of order $a$, and would become inconsequential as the size of the cavity is increased.

However, suppose we consider a spherical cavity of radius $a$, and rather than periodic boundary conditions \cite{R5}, instead require that the modes have radial wave functions which vanish at the surface of the cavity. The radial functions would then obey $j_{\ell}(\omega a)=0$, with the allowed frequencies then being given by the zeroes of the Bessel functions. Each Bessel function has its own infinite set of discrete zeroes [e.g. the zeroes of $j_0(x)=(\sin x)/x$ obey $\sin x=0$, and those of $j_1(x)=(\sin x)/x^2-(\cos x)/x$ obey $\sin x=x\cos x$]. On labeling these zeroes as $j_n^{\ell}$, inside a flat space spherical cavity the allowed frequencies are given by the discrete $\omega_n^{\ell}=j_n^{\ell}/a$, and the radial wave functions themselves are given by $j_{\ell}(j_n^{\ell} r/a)$. [In flat space there is no distinction between the radial coordinates $r$ and $\rho$ of (\ref{E2}) and (\ref{E4}).] The sum over modes proceeds initially as previously, with (\ref{E16}) being replaced by 
\begin{eqnarray}
 T_{00}(\omega_n^{\ell},r,\ell)&=&\frac{(2\ell+1)}{8\pi}\left[\left((\omega_n^{\ell})^2+
 \frac{\ell(\ell+1)}{r^2}\right)[j_{\ell}(j_n^{\ell} r/a)]^2
 +\left(\frac{dj_{\ell}(j_n^{\ell} r/a)}{dr}\right)^2\right],
 \nonumber \\
 T_{rr}(\omega_n^{\ell},r,\ell)&=&\frac{(2\ell+1)}{8\pi}\left[\left((\omega_n^{\ell})^2-
 \frac{\ell(\ell+1)}{r^2}\right)[j_{\ell}(j_n^{\ell} r/a)]^2
 +\left(\frac{dj_{\ell}(j_n^{\ell} r/a)}{dr}\right)^2\right],
 \nonumber\\
\frac{T_{\theta\theta}(\omega_n^{\ell},r,\ell)}{r^2}&=&\frac{T_{\phi\phi}(\omega_n^{\ell},r,\ell)}{r^2\sin^2\theta} =
\frac{(2\ell+1)}{8\pi}
 \left[(\omega_n^{\ell})^2[j_{\ell}(j_n^{\ell} r/a)]^2
 -\left(\frac{dj_{\ell}(j_n^{\ell} r/a)}{dr}\right)^2\right].
 \nonumber\\
 \label{E37}
\end{eqnarray}

\begin{figure}[t]
\centerline{\epsfig{file=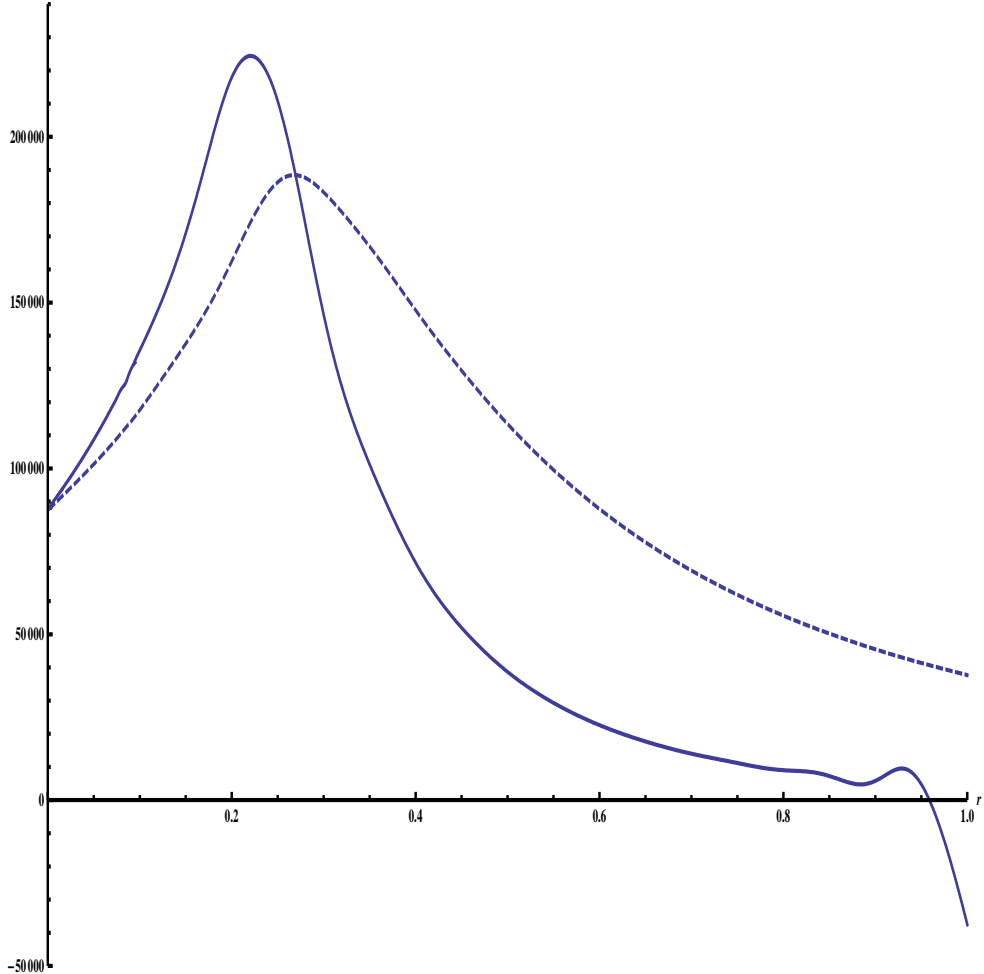,width=5.1in,height=5.1in}}
\medskip
\caption{Plot of $T_{rr}$ (dashed curve) and $T_{\theta\theta}/r^2$ (continuous curve) as a function of $r$ as incoherently summed over the first 100 zeroes of each of the first 100 $\ell$ values of $j_{\ell}(j_n^{\ell} r/a)$ in a spherical cavity of radius $a=1$.}
\label{Fig. 1}
\end{figure}

To complete the incoherent averaging we would need to sum over all $\omega_n^{\ell}$ for a fixed $\ell$ and then sum over all $\ell$. However, to see whether or not a perfect fluid form might emerge, we only need to look near $r=a$. Noting that every Bessel function obeys the recursion relation
\begin{equation}
\frac{dj_{\ell}(\omega r)}{dr}= \frac{ \ell}{r}j_{\ell}(\omega r)-\omega j_{\ell+1}(\omega r),
\label{E38}
\end{equation}
and recalling that the zeroes of $j_{\ell}(x)$ are distinct from those of $j_{\ell+1}(x)$, we see that at $r=a$
$j_{\ell}(j_n^{\ell} r/a)$ vanishes but $dj_{\ell}(j_n^{\ell} r/a)/dr$ does not. Thus at $r=a$ we have 
\begin{equation}
T_{00}(\omega_n^{\ell},a,\ell)=T_{rr}(\omega_n^{\ell},a,\ell)=-\frac{1}{a^2} T_{\theta\theta}(\omega_n^{\ell},a,\ell)=\frac{(2\ell+1)}{8\pi}
\left(\frac{dj_{\ell}(j_n^{\ell} r/a)}{dr}\right)^2\bigg{|}_{r=a}.
 \label{E39}
\end{equation}
Since $T_{rr}(\omega_n^{\ell},a,\ell)$ is positive definite and equal and opposite to the negative definite $T_{\theta\theta}(\omega_n^{\ell},a,\ell)/a^2$, and since the sum over $\omega_n^{\ell}$ and $\ell$  does not change this, we see that we do not recover the perfect fluid form at $r=a$. A straightforward Taylor series expansion of the form $r=a-\epsilon$ shows that this result remains true to second order in $\epsilon$. To get a sense of how the incoherent sum at arbitrary $r$ might look, we have numerically summed over the first 100 zeroes of each of the first 100 $\ell$ values of $j_{\ell}(j_n^{\ell} r/a)$ (i.e. 10,000 terms in total), and as we see in Fig. (1), there is no sign of a perfect fluid form.  Even in flat spacetime then, boundary conditions can prevent a fluid from being of the prefect fluid form when the fluid is in thermal equilibrium inside a spherical cavity of finite radius.

\section{Implications of Kinetic Theory}
\label{s6}

While the gravitational equations themselves provide no specific basis for leaving out any $q_f$ type terms from the fluid $T_{\mu\nu}$ in either weak or strong gravity, through the use of kinetic theory one is  able to show that such terms are not appreciable for weak gravity systems. In applying kinetic theory to gravitational systems there are two approaches, one based on the Boltzmann equation, and the other based on the Liouville equation. Since there are some differences between these two approaches \cite{R6} with the former being more appropriate for stars and the latter being more appropriate for systems such as a cluster of galaxies, we describe them both.

Consider first a set of particles each of mass $m$ in some long-range (typically gravitational) external potential $V_{\rm ext}({\bf x})$ that are undergoing rapid (typically atomic) momentum conserving two-body collisions ${\bf v}+{\bf v}_2 \rightarrow {\bf v}^{\prime}+{\bf v}_2^{\prime}$ through a scattering angle $\Omega$ with  differential collision cross section $\sigma(\Omega)$. In the absence of two-body correlations the one-particle distribution function $f({\bf x}, {\bf v},t)$ obeys the Boltzmann equation (see e.g. \cite{R7})
\begin{eqnarray} 
&&{\partial f({\bf x}, {\bf v},t) \over \partial t}+ 
{\bf v} \cdot {\partial f({\bf x}, {\bf v},t) \over \partial {\bf x}}-
{\partial V_{ext}({\bf x}) \over \partial {\bf x}} \cdot {\partial f({\bf x}, {\bf v},t)
\over \partial {\bf v}}
\nonumber\\
&&=\int  |{\bf v}-{\bf v}_2| \sigma (\Omega) 
[f({\bf x}, {\bf v},^{\prime}t)f({\bf x}, {\bf v}_2^{\prime},t)-f({\bf x}, {\bf v},t)f({\bf x}, {\bf v}_2,t)] d^3{\bf v}_2 d\Omega.
 \label{E40}
\end{eqnarray}
As such, (\ref{E40}) admits of an exact time-independent Maxwell-Boltzmann type solution
\begin{equation}
f_{\rm MB}({\bf x},{\bf p},t)=C\exp\left[-\frac{m{\bf v}^2}{2kT}-\frac{mV_{\rm ext}({\bf x})}{kT}\right],
 \label{E41}
\end{equation}
where the temperature $T$ and the coefficient $C$ are independent of ${\bf x}$. (That $f_{\rm MB}$ is an exact solution is because it makes both sides of (\ref{E40}) vanish separately.) However, since averages with this distribution function are given by $\langle A \rangle=\int d^3{\bf p}Af_{\rm MB}/\int d^3{\bf p} f_{\rm MB}$, for any dynamical variable $A({\bf x}, {\bf v},t)$, the $V_{\rm ext}({\bf x})$ term would drop out and all averages evaluated with this $f_{\rm MB}$ would be spatially independent. Moreover, the number density itself would behave as $n({\bf x})=\int d^3{\bf p} f_{\rm MB}({\bf x},{\bf p},t)\sim \exp(-mV_{\rm ext}({\bf x})/kT)$, and give a spatial dependence that is nothing like that required of the density of a star of finite size. 

Now as such, use of the above Boltzmann equation presupposes that the only collisions of relevance are atomic type ones, and that those collisions dominate the relaxation of the system to the distribution function $f_{\rm MB}({\bf x},{\bf p},t)$. However, in stars gravitational collisions also play a role, and their effect cannot be isolated solely in the $\partial V_{\rm ext}({\bf x})/\partial {\bf x}$ term in (\ref{E40}). Rather, they serve to modify the right-hand side of (\ref{E40}) as well. However, since the cross-section for gravitational scattering is infinite, the effect of gravity cannot be accounted for by a collision integral term in which one simply includes a gravitational contribution to $\sigma(\Omega)$. Rather, one should work with the Liouville equation as it is better suited to handle long range forces, and we will discuss this below.

However, before doing so, we note that in kinetic theory, through use of the method of the most probable distribution, it is possible to determine the  distribution function without needing to actually solve or even construct the Boltzmann equation at all. This method does not determine the approach to equilibrium (i.e. the temporal behavior of the distribution function), but does give the equilibrium configuration that results, and it is valid even if the distribution function does not obey an equation such as (\ref{E40}) at all. Thus we can use the method of the most probable distribution to get a sense of how gravitational interactions might modify the distribution function (\ref{E41}). Ordinarily one applies the method by taking the system of interest to be confined to a region of phase space with a fixed total energy and fixed volume and introduces spatially-independent Lagrange multipliers for the total number of particles and the total energy \cite{R7}. In the absence of gravity this leads to the Maxwell-Boltzmann distribution being the overwhelmingly likely one. In the presence of non-relativistic gravity we can adapt this method by decomposing the star into concentric shells, and take the temperature and density to be a constant within any given  shell but to vary from one shell to the next. The approximation here is that particles stay within shells and do not exchange energy or momentum with any particles except those in their own shell. (Since gravity is a long range force, at best this assumption could only be valid when gravity is weak.) In this approximation the Lagrange multipliers for the total number of particles and the total energy within a given shell are taken to be uniform throughout the shell but to depend on the location of the shell, with the most probable (${\rm mp}$) distribution function throughout the star then being given by 
\begin{equation}
f_{\rm mp}({\bf x},{\bf p},t)=\frac{n}{(2\pi m \theta)^{3/2}}\exp\left[-\frac{m({\bf v}-{\bf u})^2}{2\theta}\right].
 \label{E42}
\end{equation}
In (\ref{E42}) the particle number density $n({\bf x},t)=\int d^3{\bf p} f_{\rm mp}$, the average velocity ${\bf u}({\bf x},t)=\int d^3{\bf p}f_{\rm mp}{\bf p}/mn({\bf x},t)$ and the temperature $\theta({\bf x},t)=(m/3)\int d^3{\bf p} f_{\rm mp}|{\bf v}-u({\bf x},t)|^2/n({\bf x},t)$ are now all spatially dependent.

While such a distribution function would not be an exact solution to the Boltzmann equation, in the event that $n({\bf x},t)$, ${\bf u}({\bf x},t)$ and $\theta({\bf x},t)$ are all slowly spatially varying, $f_{\rm mp}({\bf x},{\bf p},t)$ would be a good first-order approximation to it. For such a distribution function the pressure tensor $P_{ij}=mn({\bf x},t)\langle (v_i-u_i)(v_j-u_j)\rangle$ evaluates to the isotropic
\begin{equation}
P_{ij}=\delta_{ij}n({\bf x},t)\theta({\bf x},t),
 \label{E43}
\end{equation}
to thus be of none other than of perfect fluid form. If the full distribution function is taken to obey the Boltzmann equation, then in the next order the pressure tensor evaluates \cite{R7} to the anisotropic 
\begin{equation}
P_{ij}=\delta_{ij}n\theta-\tau n \theta\left[\frac{\partial u_i}{\partial x_j}+\frac{\partial u_j}{\partial x_i}-\frac{2}{3}\delta_{ij}{\bf \nabla}\cdot {\bf u}\right]
 \label{E44}
\end{equation}
where $\tau$ is the mean free time between particle collisions. With $n({\bf x},t)$ depending on position, the $\partial_ju_i+\partial_iu_j$ term would not be proportional to $\delta_{ij}$. When $n({\bf x},t)$ is slowly varying then, the pressure only begins to depart from $\delta_{ij}$ in corrections to first order. To the extent then that weak gravity would cause a star to bind with a slowly varying density and that $\tau$ would be small (i.e. a small mean free time between atomic collisions), the $q_f$ type term could be neglected in lowest order. Thus our ability to ignore $q_f$ in a weak gravity star depends on how good a dynamical approximation the above $f_{\rm mp}({\bf x},{\bf p},t)$ is to the full $f({\bf x},{\bf p},t)$.  The use of perfect fluids as gravitational sources for weak gravity stars is thus equivalent to using $f_{\rm mp}({\bf x},{\bf p},t)$ as the one-particle distribution function, and is justifiable if $n({\bf x},t)$ varies slowly enough. However, once one has to go to a more rapidly varying $n({\bf x},t)$ as would be the case for stronger gravity, there would no longer appear to be any immediate justification for ignoring $q_f$ type terms \cite{R7a}.

In terms of the geodesic equation discussion given earlier, we now see that short wavelength for eikonal purposes means short with respect to the distance scale on which the particle number density $n({\bf x},t)$ varies. The slower the spatial variation of $n({\bf x},t)$ then, the fewer the number of modes that will not incoherently average to a perfect fluid \cite{Q4}. 

In the above discussion it is the non-gravitational collisions which dominate, with the discussion being given for a weak gravity star in which there are atomic collisions between the atoms in the star. However, for a cluster of galaxies, it is gravity itself which has to establish an equilibrium distribution of galaxies. In clusters there is a lot of X-ray producing plasma located in the region between the individual galaxies in the cluster. Collisions between the atomic particles in the plasma can readily bring the plasma to thermal equilibrium, but unless they can bring the galaxy distribution to equilibrium too, it would be up to gravity  to do so. 

To describe the role that gravity plays in a cluster of galaxies we turn to the Liouville equation. We treat each of the $N$ galaxies in the cluster as a non-relativistic point particle of mass $m$ with position ${\bf x}_i$ and velocity ${\bf v}_i$. Each galaxy moves in the non-relativistic  gravitational field $\phi({\bf x}_i)$ produced by the other galaxies in the cluster, and obeys
\begin{equation}
\frac{d^2{\bf x}_i}{dt^2} = -\frac{\partial \phi({\bf x}_i)}{\partial {\bf x}_i}. 
 \label{E45}
\end{equation}
One introduces the normalized (to one) $6N$-dimensional distribution function $f^{(N)}({\bf x}_1,{\bf v}_1,.....,{\bf x}_N,{\bf v}_N,t)$, and finds that it obeys the Liouville equation
\begin{equation}
\frac{df^{(N)}}{dt}=\frac{\partial f^{(N)}} { \partial t}+\sum_{\alpha=1}^{N} \left[ 
{\bf v}_{\alpha} \cdot \frac{\partial f^{(N)}}{\partial {\bf x}_{\alpha}}-
\frac{\partial \phi_{\alpha}}{\partial {\bf x}_{\alpha}} \cdot\frac{\partial f^{(N)} }{ \partial 
{\bf v}_{\alpha}} \right]=0, 
 \label{E46}
\end{equation}
where 
\begin{equation}
\phi_{\alpha}({\bf x}_\alpha)=\sum_{\beta \neq \alpha}\phi({\bf x}_\alpha, {\bf x}_\beta).
 \label{E47}
\end{equation}
If the distribution function is symmetric under the interchange of any pair of particles and sufficiently convergent asymptotically, upon integrating (\ref{E46}), one finds (see e.g. \cite{R8}) that for large $N$, the one- and two-particle distribution functions
\begin{eqnarray}
f^{(1)}({\bf x}_1,{\bf v}_1,t)&=&\int d^3{\bf x}_2d^3{\bf v}_2....d^3{\bf x}_Nd^3{\bf v}_Nf^{(N)},
\nonumber\\
f^{(2)}({\bf x}_1,{\bf v}_1,{\bf x}_2,{\bf v}_2,t)&=&\int d^3{\bf x}_3d^3{\bf v}_3....d^3{\bf x}_Nd^3{\bf v}_Nf^{(N)},
 \label{E48}
\end{eqnarray}
are related by 
\begin{equation}
\frac{\partial f^{(1)}} { \partial t}+
{\bf v}_{1} \cdot \frac{\partial f^{(1)}}{\partial {\bf x}_{1}}=N\int  \frac{\partial \phi({\bf x}_1,{\bf x}_2)}{ \partial {\bf x}_{1}} \cdot 
\frac{\partial f^{(2)}}{ \partial{\bf v}_{1}}d^3{\bf x}_2d^3{\bf v}_2.
 \label{E49}
\end{equation}
In terms of the two-body correlation function defined by
\begin{equation}
g({\bf x}_1,{\bf v}_1,{\bf x}_2,{\bf v}_2,t)=f^{(2)}({\bf x}_1,{\bf v}_1,{\bf x}_2,{\bf v}_2,t)-
f^{(1)}({\bf x}_1,{\bf v}_1,t)f^{(1)}({\bf x}_2,{\bf v}_2,t)
 \label{E50}
\end{equation}
and the kinetic theory distribution $f({\bf x}_1,{\bf v}_1,t)=Nf^{(1)}({\bf x}_1,{\bf v}_1,t)$ that is normalized to 
\begin{equation}
\int d^3{\bf v}f({\bf x},{\bf v},t)=n({\bf x},t),\qquad \int d^3{\bf x}n({\bf x},t)=N,
 \label{E51}
\end{equation}
we find that (\ref{E49}) takes the form 
\begin{eqnarray}
&&\frac{\partial f({\bf x},{\bf v},t)}{\partial t}+ 
{\bf v} \cdot \frac{\partial f({\bf x},{\bf v},t) }{ \partial {\bf x}}-
\frac{\partial V({\bf x}) }{ \partial {\bf x}} \cdot \frac{\partial f({\bf x},{\bf v},t) 
}{ \partial{\bf v}}
\nonumber\\
&&=N^2 \frac{\partial }{\partial {\bf v}} \cdot 
\int {\partial \phi({\bf x},{\bf x}_{2}) \over \partial {\bf x}}
g({\bf x},{\bf v},{\bf x}_2,{\bf v}_2,t)d^3{\bf x}_2d^3{\bf v}_2
 \label{E52}
\end{eqnarray}
where
\begin{equation}
V({\bf x})=\int d^3{\bf x}_2d^3{\bf v}_2f({\bf x}_2,{\bf v}_2,t)\phi({\bf x},{\bf x}_2)
=\int d^3{\bf x}_2n({\bf x}_2,t)\phi({\bf x},{\bf x}_2).
\label{E53}
\end{equation}

As such, the potential $V({\bf x})$ introduced in (\ref{E53}) serves as a mean-field potential, and should the system relax to a steady state in which  the two-body correlation function becomes negligible, the left-hand side of (\ref{E52}) would then become equal to zero. At first glance the left-hand side of (\ref{E52}) looks quite like the left-hand side of the Boltzmann equation (\ref{E40}). However, in (\ref{E40}) $V_{\rm ext}({\bf x})$ is an external one-body potential, while in (\ref{E52}) $V({\bf x})$ is a statistically averaged two-body potential. Moreover, while the vanishing of $g({\bf x}_1,{\bf v}_1,{\bf x}_2,{\bf v}_2,t)$ would cause the right-hand side of (\ref{E52}) to vanish, it would not have the same effect on the right-hand side of the Boltzmann equation of (\ref{E40}). In fact, it was the very requirement that $g({\bf x}_1,{\bf v}_1,{\bf x}_2,{\bf v}_2,t)$ vanish that led \cite{R7} to the explicit form for the collision integral term given on the right-hand side of (\ref{E40}) in the first place, with the vanishing of $g({\bf x}_1,{\bf v}_1,{\bf x}_2,{\bf v}_2,t)$ not requiring the vanishing of the Boltzmann equation collision integral term. With regard to (\ref{E52}), we note that when the right-hand side of (\ref{E52}) does vanish (i.e. when the system virializes \cite{R9}), one can say only that in the steady state solution to (\ref{E52}) the virialized $f({\bf x},{\bf v},t)$ has to be a function of the quantity ${\bf v}^2/2+V({\bf x})$ (and of the angular momentum ${\bf L}={\bf x}\times {\bf v}$ as well if the system is rotating). With (\ref{E52}) possessing no analog of the collision integral term in (\ref{E40}) (gravity being long range), there is nothing to force $f({\bf x},{\bf v},t)$ to be an exponential function of ${\bf v}^2/2+V({\bf x})$ \cite{R10}. The specific dependence on ${\bf v}^2/2+V({\bf x})$ that a virialized $f({\bf x},{\bf v},t)$ would acquire would depend entirely on how the two-body  $g({\bf x}_1,{\bf v}_1,{\bf x}_2,{\bf v}_2,t)$  would behave before it becomes negligible. However, the very fact that the virialized $f({\bf x},{\bf v},t)$ does only depend on the magnitude of ${\bf v}$ and not on its direction (i.e. not on ${\bf L}$) then enables us to recover the isotropic structure of (\ref{E43}), with the pressure tensor of a steady state weak gravity cluster of galaxies indeed being of the perfect fluid form, just as desired. However, the situation that is to obtain in the  general relativistic strong gravity  case remains to be explored. 

PDM would like to thank Dr.~ Klaus~Kirsten for helpful correspondence.

{}

\begin{thebibliography}{}

\bibitem{R3a} G.~F.~R.~Ellis,~{\it Relativistic Cosmology}, in Proceedings of the International School of Physics ``Enrico Fermi", Course XLVII, 1969, B.~K.~Sachs, Editor, Academic Press, New~York,~N.~Y.~(1971).

\bibitem{R4} P.~D.~Mannheim,~ Prog.~Part.~Nucl.~Phys.~{\bf 56},~340~(2006). 

\bibitem{Q1} This is not all that can happen when one goes to curved space, since the wave equation of the scalar field itself can change. For instance, general covariance does not forbid the presence in the action of a direct coupling term $(\xi/12)S^2R^{\alpha}_{\phantom{\alpha}\alpha}$ between the scalar field and the Ricci scalar, with the scalar field equation of motion then being modified into $\nabla^{\mu}\nabla_{\mu}S+(\xi/6) SR^{\alpha}_{\phantom{\alpha}\alpha}=0$ where $\xi$ is a constant.


\bibitem{Q2} S. Weinberg, {\it Gravitation and Cosmology:
Principles  and Applications of the General Theory of Relativity} 
(Wiley, New York, 1972)




\bibitem{Q3} The equivalence principle does not require that all effects associated with a gravitational field can be removed at a given point by a coordinate transformation, as any Riemann tensor dependent term can never be brought to zero by a coordinate transformation. Rather, the content of the equivalence principle is  that at any given point in a curved space the Christoffel symbol dependent contribution to the geodesic equation can be removed via a coordinate transformation. Specifically, since the Christoffel symbols are not coordinate tensors, at any given point they can be brought to zero via a general coordinate transformation. Similarly, the quantity $d^2x^{\lambda}/d\tau^2$ is not a coordinate tensor either. It is only the specific linear combination $d^2x^{\lambda}/d\tau^2+\Gamma^{\lambda}_{\mu\nu}(dx^{\mu}/d\tau)( dx^{\nu}/d\tau)$ with the two terms having this very specific relative weight that is a coordinate tensor, to thus enforce the equality of the gravitational and inertial masses in (\ref{D10}) and (\ref{D11}), even as these two equations contain Riemann tensor dependent terms.

\bibitem{R1} P.~D.~Mannheim~and~D.~Kazanas,~Gen.~Rel.~Gravit.~{\bf 20},~201~(1988).

\bibitem{R2} Y.~Deng~and~P.~D.~Mannheim,~Gen.~Rel.~Gravit.~{\bf 20},~969~(1988).

\bibitem{R3} Y.~Deng~and~P.~D.~Mannheim,~Astrophys.~Sp.~Sci.~ {\bf 135},~261~(1987).

\bibitem{R3aa} J.~ Ehlers,~Akad.~Wiss.~Lit. ~Mainz,~Abhandl.~Math.-Nat.~Kl.,~{\bf 11}, 792 (1961), 
 [English translation: Gen.~Rel.~Grav.~ {\bf 25}, 1225 (1993)].

\bibitem{R3b} In passing we note that while the energy-momentum tensor given in (\ref{E29a}) and (\ref{E30}) recovers the perfect fluid form when one takes its flat space limit (as it of course must), it does so not by having $\pi_{\mu\nu}$ be a geometric quantity that vanishes in flat space (as would be the case if $\pi_{\mu\nu}$ were, say,  to be built out of tensors constructed from the Riemann tensor), but rather by having its coefficient $q_f$ vanish in the limit. In the flat space limit the $\pi_{\mu\nu}$ term thus vanishes dynamically rather than kinematically.


\bibitem{R4a} With the vectors $n^1_{\mu}=U_{\mu}+\sin\alpha V_{\mu}$ and $n^2_{\mu}=U_{\mu}+\sin\beta W_{\mu}$ being timelike for $U_{\mu}=(H^{1/2},0,0,0)$, $V_{\mu}=(0,J^{1/2},0,0)$, $W_{\mu}=(0,0,\rho J^{1/2},0)$ and arbitrary angles $\alpha$ and $\beta$, and with $n^1_{\mu}n^1_{\nu}T^{\mu\nu}$ and $n^2_{\mu}n^2_{\nu}T^{\mu\nu}$ respectively evaluating to $\rho_f+\sin^2\alpha (p_f+2q_f)$ and $\rho_f+\sin^2\beta(p_f-q_f)$, for large enough positive or negative $q_f$, it might be possible to violate the weak energy condition $n_{\mu}n_{\nu}T^{\mu\nu} >0$ for some appropriate timelike $n_{\mu}$ and some specific set of $\rho_f$, $p_f$ and $q_f$. (This happens not to be the case for the particular $\rho_f$, $p_f$ and $q_f$ given in (\ref{E30}) and the metric given in (\ref{E19}).) For strong gravity systems then, by generating the appropriate $q_f$ term, it might be possible to evade the Hawking-Penrose singularity theorems for collapsing stars by relaxing one of the assumptions which goes into the proof. Moreover, we would note that since a condition such as the weak energy condition is initially motivated by familiarity with standard fluids in flat spacetime (where there is no $q_f$ term and the quantity $\rho_f+p_f$ is positive), its extension to curved space presupposes that the only role of gravity is to generalize a flat space condition to its covariant form, and not to act dynamically in a way that might prevent the covariant generalization from actually holding.

\bibitem{R5} For modes of the form $j_{\ell}(\omega r)P^m_{\ell}(\theta)$, periodic boundary conditions require that $\ell$ be even.

\bibitem{R6} P.~D.~Mannheim,~{\it Linear potentials in galaxies and clusters of galaxies},~astro-ph/9504022,~April~1995. 

\bibitem{R7} K.~Huang~{\it Statistical Mechanics},~Second~Edition,~J.~Wiley,~New~York,~N.~Y.~(1987).

\bibitem{R7a} In the theory of white dwarf stars the degeneracy pressure of the electrons in the star is ordinarily calculated by introducing plane wave momentum eigenstates and filling up the Fermi sea to a maximum momentum $k_F=\hbar(3\pi^2 n)^{1/3}$, where $n$ is the electron number density.  Since such momentum eigenstates are associated with spatial translation invariance, in order to be able to use them at all the number density $n$ and the pressure $p=(1/3\pi^2\hbar^3)\int_0^{k_F}dk k^4/(k^2+m_e^2)^{1/2}$ would have to be independent of position,  and thus not be able to provide the pressure gradient needed to balance the gravitational attraction of the nuclei in the star. To obtain the needed pressure gradient in a weak gravity star one must use not the equilibrium Maxwell-Boltzmann (or Fermi-Dirac) distribution $f_{\rm MB}({\bf x}, {\bf p},t)$ but rather a spatially dependent departure from it, viz. the most probable distribution $f_{\rm mp}({\bf x}, {\bf p},t)$ as evaluated with slowly varying number density, average velocity, and temperature. One thus considers the star to be divided into concentric shells, with there being no spatial variation within any specific shell (so that one can use plane wave momentum states within each such shell), but with each shell having a maximum momentum $k_F$ that depends on the radius of the shell. In this way the pressure becomes dependent on the radial dependence of the number density $n$, and thereby generates the needed pressure gradient; and one is thus able to use momentum eigenstates in a weak gravity star even though the electron number density depends on position. However, for a strong gravity star one is no longer free to use momentum eigenstates and restrict to slowly varying modifications to the Maxwell-Boltzmann distribution $f_{\rm MB}({\bf x}, {\bf p},t)$. Rather, to construct the partition function, one must explicitly solve the Dirac equation in the curved space background and use whatever solutions to the radial equation then emerge. As we noted in Sec.~\ref{s3}, the full calculation is a highly non-linear one in which one must construct the energy-momentum tensor by an incoherent averaging over the solutions to the curved space wave equation, and then have this very same energy-momentum tensor  serve as the source of the Einstein equations so as to fix the metric coefficients that are to be used to obtain the solutions to the curved space wave equation in the first place. 

\bibitem{Q4} For a normal star (viz. one far from any possible cold late time state) the interior temperature is quite high. With high temperatures favoring high frequency Boltzmann factors, again we see that in normal stars low frequency modes are suppressed.


\bibitem{R8} J.~Binney~and~S.~Tremaine,~{\it Galactic Dynamics},~Princeton~University~Press,~Princeton,~N.~J.~ (1987).



\bibitem{R9} As noted in \cite{R8}, by multiplying (\ref{E52}) by ${\bf x}\cdot{\bf v}$, integrating over all ${\bf v}$ and all ${\bf x}$, and dropping all asymptotic surface terms, one obtains $\int d^3{\bf x}\langle {\bf v}^2 \rangle-\int d^3{\bf x}\langle {\bf x}\cdot \partial V({\bf x})/\partial {\bf x}\rangle=\partial(\int d^3{\bf x} \langle{\bf x}\cdot{\bf v}\rangle)/\partial t+N^2\int d^3{\bf x}d^3{\bf v}d^3{\bf x}_2d^3{\bf v}_2g({\bf x},{\bf v},{\bf x}_2,{\bf v}_2,t){\bf x}\cdot\partial\phi({\bf x},{\bf x}_2)/\partial {\bf x}$, an expression which reduces to the familiar virial relation when there are no correlations.

\bibitem{R10} The use of the method of the most probable distribution as used above for stars is not readily applicable here, since for particles whose motions are controlled by long range forces alone (i.e. in the absence of short range forces), the cluster cannot readily be approximated by shells of particles that do not  exchange energy and momentum with particles in other shells.



\end{thebibliography}
\end{document}